\documentclass{emulateapj}
\usepackage{url}
\usepackage{graphicx}
\usepackage{amssymb}
\usepackage{amsmath}

\shorttitle{Forming chondrules in impact splashes}
\shortauthors{Dullemond, Stammler, Johansen}

\begin{document}

\title{Forming chondrules in impact splashes\\ I. Radiative cooling model}

\author{Cornelis Petrus Dullemond, Sebastian Markus Stammler}
\affil{Institute for Theoretical Astrophysics, Heidelberg University,
Albert-Ueberle-Strasse 2, 69120 Heidelberg, Germany}

\author{Anders Johansen}
\affil{Lund Observatory, Department of Astronomy and Theoretical
Physics, Lund University, Box 43, 22100 Lund, Sweden}

\submitted{The Astrophysical Journal, 794:91 (12pp), 2014 October 10}

\begin{abstract}
  The formation of chondrules is one of the oldest unsolved mysteries in
  meteoritics and planet formation. Recently an old idea has been revived:
  the idea that chondrules form as a result of collisions between
  planetesimals in which the ejected molten material forms small droplets
  which solidify to become chondrules. Pre-melting of the planetesimals by
  radioactive decay of $^{26}$Al would help producing sprays of melt even at
  relatively low impact velocity. In this paper we study the radiative
  cooling of a ballistically expanding spherical cloud of chondrule droplets
  ejected from the impact site. We present results from a numerical
  radiative transfer models as well as analytic approximate solutions. We
  find that the temperature after the start of the expansion of the cloud
  remains constant for a time $t_{\mathrm{cool}}$ and then drops with time
  $t$ approximately as $T\simeq T_0[(3/5)t/t_{\mathrm{cool}}+ 2/5]^{-5/3}$
  for $t>t_{\mathrm{cool}}$. The time at which this temperature drop starts
  $t_{\mathrm{cool}}$ depends via an analytical formula on the mass of the
  cloud, the expansion velocity and the size of the chondrule. During
  the early isothermal expansion phase the density is still so high that we
  expect the vapor of volatile elements to saturate so that no large
  volatile losses are expected.
\end{abstract}


\section{Introduction}
\label{sec-introduction}
The formation of chondrules is one of the fundamental questions of
meteoritics and planet formation. Chondrules are the 0.1$\cdots$1
mm-size once-molten silicate spherules found abundantly in primitive
meteorites known as chondrites (Jones, Grossman \& Rubin 2005; Sears 2004;
Davis et al.~2014). Most of these chondrites in fact consist predominantly
of these chondrules, so the melt-producing events that created them must
have been extremely common during the first few million years of the solar
system. Yet there is confusing and conflicting evidence as to what these
events might have been. Boss (1996) published an overview of the status of
the discussion at that time, though significant new developments have
occurred since then.

Many theories have been put forward over the last half a century. One theory
involves impact melt sprays. Put forward by Urey (1953) and refined by
Kieffer (1975) this model states that high-velocity impacts ($\gtrsim$
3$\cdots$5 km/s) could lead to sprays of impact melt that produce droplets
which solidify into chondrules. While these required collision velocities
are high, a small fraction of impacts might acquire such velocities (Bottke
et al.~1994). Perhaps the so-called ``Grand Tack''-scenario of Walsh et
al.~(2011), in which Jupiter temporarily entered the asteroid belt region
before migrating back outward, might produce such high velocities. However,
Taylor et al.~(1983) put forward a number of arguments against the
impact-melt scenario, some of which were based the differences between
chondrites and lunar impact regolith. 

Zook (1980) suggested that if the interiors of the colliding bodies are
already in a molten state due to $^{26}$Al decay, then the required impact
velocities to create sprays of melt are much lower, and thus more consistent
with expectations of the average relative velocities between
planetesimals. Sanders \& Taylor (2005), Hevey \& Sanders (2006) and Sanders
\& Scott (2012) follow up on this idea of pre-molten impactors and back it
up with models and meteoritic evidence. Sanders \& Scott (2012) argue that
this scenario is hard to avoid, given that in the first 2.5 million years
most planetesimals were internally nearly fully molten by $^{26}$Al decay
heat, and that collisions between planetesimals were extremely common. Each
collision would almost certainly release substantial amounts of molten rock
into the nebula in the form of sprays of lava droplets, and it is natural to
assume that these may be chondrules. They argue that the near-solar
composition of chondrules can be explained by the vigorous convection inside
the molten planetesimals that may slow down iron/nickel-core formation, thus
keeping the melt solar. Some degree of differentiation would then in fact
explain the low iron-content L and LL chondrites.

Recently, Asphaug et al.~(2011) performed Smooth Particle Hydrodynamics
simulations for such impacts to demonstrate the dynamics of this
scenario. In addition they proposed a simple way to calculate the melt
droplet size by equating the released enthalpy after the collision to the
surface energy of the droplets.

There are many alternative scenarios proposed in the literature. Perhaps the
most popular model is the flash heating by nebular shocks (Hood \&
Hor\'anyi, 1991). Desch \& Connolly (2002) and Ciesla \& Hood (2002)
developed detailed 1-D radiative shock models with dust particles and
chondrules interacting both radiatively and frictionally with the gas. They
showed how the radiative shocks exhibit temperature spikes of mere tens of
seconds (with cooling rates $>10^4$ K/hr) that would be good candidates
for chondrule-forming events. After the main shock temperature spike their
model exhibits a further cooling at intermediate cooling rate ($\sim$50
K/hr), which cools the chondrules to sufficiently low temperature in a
sufficiently short amount of time. This model seems to produce the
flash-heating events required for turning ``dust bunnies'' into
chondrules. 
However, so far the shock scenario still has several unresolved issues
  (e.g., Desch et al. 2012; Boley et al. 2013; Stammler \& Dullemond 2014).

Also some issues are raised about whether large scale
shocks in the optically thick solar nebula can explain the time scales
involved in chondrule formation (Stammler \& Dullemond 2014). It is
noteworthy that there exist several other nebular flash heating models, most
notably flash heating by nebular lightning (Gibbard et al.~1997) and flash
heating by energy dissipation in current sheets forming in MHD turbulence
(Hubbard et al.~2012).

One important constraint on chondrule formation models is that these heating
events were very short-lived compared to any other nebular time scales. By
comparing textures of chondrule to those obtained in furnace experiments it
can be inferred that chondrules must have cooled from the liquidus
temperature down to below the solidus temperature in a matter of hours
(Hewins 1983; Hewins \& Connolly 1996; see also references in Morris \&
Desch 2010 and the excellent review paper of Desch et al.~2012). In
other words: the chondrule forming events must have been flash-events. On
the other hand, under optically thin conditions a molten chondrule would
radiatively cool in about a second, which would be too fast.

Another constraint that a chondrule formation model must fulfill is the
retention of volatile elements such as Fe, Mg, Si, Na and K. Chondrules are
not observed to have low abundances of these elements. In particular for the
highly volatile elements Na and K this is a puzzle, because any heating
event that heats a (proto-)chondrule above $\sim$ 1700 K and keeps these
chondrules above that temperature for more than a few minutes will cause
most of the Na and K to evaporate out of the chondrule. Alexander et
al.~(2008) argue that this means that chondrules must have formed in regions
that are extremely dense in solids, much more so that typical dust
concentration mechanisms in the protoplanetary disk can achieve. According
to their calculations such regions are so dense in solids that they must be
self-gravitating. Morris et al.~(2012) instead propose that chondrules
  formed behind the bow shock of a fast moving planetary embryo, and that
  the embryo's atmosphere is rich in such volatile elements caused by
  outgassing from the embryo's interior, thus providing the necessary
  vapor pressure to keep the chondrules volatile-rich.

In this paper we will focus on the impact splash hypothesis, either the
low-velocity pre-molten planetesimal version or the high-velocity
impact-melt version. We assume that after the collision between two
planetesimals a cloud of molten droplets of magma was released. Since soon
after the ejection of this cloud of magma droplets the internal pressure of
the cloud would have dropped to near-zero (there is, to good approximation,
only vacuum between the droplets), the cloud would simply expand
ballistically. Initially the density of the cloud is so high that the cloud
is completely optically thick and no radiation can escape from its
interior. There is also no adiabatic cooling because of the lack of
pressure.  So the temperature of the magma droplets will initially stay
roughly constant in time. As the cloud expands, however, the optical depth
drops and eventually the cloud will start to cool radiatively. During the
early stages the density is very high and the evaporation of volatile
elements will be saturated (as was argued by Alexander et al.~2008). Once
the temperature drops below the solidus, volatiles can no longer escape.

In this first paper we intend to compute how the temperature behaves as a
function of time after the impact. To do this we set up a simple model: that
of a homologously ballistically expanding homogeneous spherical cloud of
lava droplets. We compute the temperature of the chondrules as a function of
time $t$ and comoving radial location inside the cloud
$r/r_{\mathrm{cloud}}$ by solving the time-dependent radiative transfer
equation. We also present an analytic approximate solution. While the
spherical cloud model is not an accurate model of the complex shape of an
impact splash, it can be regarded as a model of {\em part} of the impact
splash. As such we believe that it will give reasonable estimates of the
radiative cooling behavior of the impact splash as a function of model
parameters such as the total mass of the cloud and its expansion
velocity. The total mass of the cloud tells something about the masses of
the two colliding bodies while the expansion velocity of the cloud tells us
something about the impact velocity.

\section{Expanding cloud model}
\label{sec-cloud-model}
\begin{figure*}
\begin{center}
\includegraphics[width=0.9\textwidth]{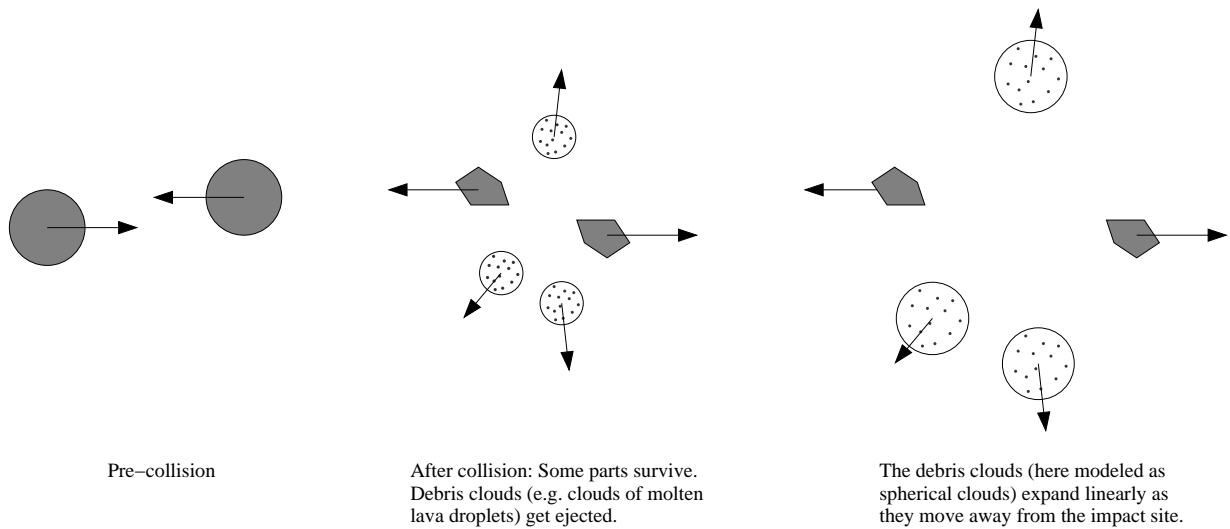}
\end{center}
\caption{\label{fig-pictogram}Pictographic representation of the model.
Left: Two planetesimals approach each other and are about to collide.
Middle: After the collision some parts of the planetesimals may survive,
but some parts are destroyed and dispersed in a cloud of debris. We model
this dispersing cloud as a set of spherical clouds. If the debris consists
of melt, the cloud will consist of molten droplets (chondrules). Right:
As the clouds move away from the point of impact they expand linearly.
The average distance between the chondrules increases linearly with
time.}
\end{figure*}

When lava droplets are produced in a collision between two planetesimals,
they will disperse away from the impact site in a ballistic way. Let us call
this the ``impact splash''. If the impact velocity is larger than the escape
velocity of the two colliding planetesimals this ballistic (pressureless)
expansion will be linear (i.e.\ the velocity of expansion will not change
with time). We regard the impact splash as an expanding cloud of lava
droplets (chondrules-to-be). This cloud is not necessarily centered on the
impact site; it can also move away from the impact site. The radius of the
cloud increases linearly with time $t$ and thus the density of the cloud
will drop as $1/t^3$. Depending on the complexity of the geometry of the
impact splash (see e.g. Asphaug et al.~2011) we can also consider the splash
to be multiple smaller clouds.  In either case, each of these clouds will
expand linearly. The scenario is pictographically shown in
Fig.~\ref{fig-pictogram}.

If the impact is at low velocity, then the gravitational
pull of the surviving (or merged) body can cause some (or most) of the
material to re-accrete. In this case the $1/t^3$ expansion will cease and
turn into recompression. Initially, however, the linear expansion and
$1/t^3$ density drop will still be a good description.

To compute the temperature of the droplets in these clouds as a function of
time after impact we must solve the time-dependent equation of radiative
transfer. This is a very non-local problem, because a cooling droplet at
location $\vec x_1$ radiates away some heat into the form of infrared
radiation, which can then be absorbed by another droplet at location $\vec
x_2\neq \vec x_1$ which is then heated. The droplets are thus radiatively
coupled over large distances. Solving this problem in 3-D for complex cloud
geometries is somewhat challenging. 

We believe, however, that much can already be learned from a simple
spherical cloud model, for which the radiative transfer problem can be
solved to high precision and reliability with the 1-D tangent-ray variable
eddington factor method. We will describe the method in
Section \ref{sec-radiative-transfer}. 

The three main parameters of the model are:
\begin{equation}
M_{\mathrm{cloud}} \qquad , \qquad T_0 \qquad , \qquad v_{\mathrm{exp}}
\end{equation}
where $M_{\mathrm{cloud}}$ is the total mass worth of chondrules (lava
droplets) in our spherical cloud, $T_0$ is the initial temperature of the
chondrules and $v_{\mathrm{exp}}$ is the expansion velocity of the outer
radius of the cloud. To get a feeling for the numbers it is more
convenient to express the mass of the cloud in terms of the radius 
$R_{\mathrm{melt,0}}$ of a ball of magma of the same mass $M_{\mathrm{cloud}}$:
\begin{equation}\label{eq-def-rmelt0}
M_{\mathrm{cloud}}=\frac{4\pi}{3}\xi_{\mathrm{chon}}R_{\mathrm{melt,0}}^3
\end{equation}
where $\xi_{\mathrm{chon}}$ is the material density of the chondrule
droplets and thus, by definition, the material density of the hypothetical
ball of magma. For a linearly expanding cloud the radius of the cloud is
\begin{equation}\label{eq-rcloud-afo-t}
R_{\mathrm{cloud}}(t) = v_{\mathrm{exp}}t
\end{equation}
where the time $t$ is the time since the impact and the ejection of the
cloud of melt droplets, assuming perfectly ballistic (pressureless)
expansion and of course assuming large enough $t$ that
$R_{\mathrm{cloud}}(t)\gg R_{\mathrm{melt,0}}$ so that the droplets are
clearly separated from each other. The velocity profile inside the cloud
is
\begin{equation}
v(r,t) = v_{\mathrm{exp}}\frac{r}{R_{\mathrm{cloud}}(t)}
\end{equation}
which is of course only valid for $r\le R_{\mathrm{cloud}}(t)$.

Let us define a coordinate $r$ centered on the center of the spherical
cloud.  We assume that the density within the cloud is constant and outside
of the cloud is zero. The density of the cloud as a function of time is then
\begin{equation}
\rho_{\mathrm{cloud}}(r,t) = \left\{
\begin{matrix}
\frac{3M_{\mathrm{cloud}}}{4\pi R_{\mathrm{cloud}}^3(t)} & \qquad & \hbox{for}\; r\le R_{\mathrm{cloud}}(t)\\
0 & \qquad & \hbox{for}\; r> R_{\mathrm{cloud}}(t)
\end{matrix}
\right.
\end{equation}
From here on we write only $\rho_{\mathrm{cloud}}(t)$ instead of
$\rho_{\mathrm{cloud}}(r,t)$. The time-dependence of $\rho_{\mathrm{cloud}}(t)$  is
\begin{equation}\label{eq-rho-afo-time}
\rho_{\mathrm{cloud}}(t) = \frac{3M_{\mathrm{cloud}}}{4\pi v_{\mathrm{exp}}^3}\frac{1}{t^3}
\end{equation}
Let us consider chondrules (lava droplets) of radius $a_{\mathrm{chon}}$ and
material density $\xi_{\mathrm{chon}}$. We take $\xi_{\mathrm{chon}}=3.3$
g/cm$^3$ for our model. The mass of a chondrule is then
\begin{equation}\label{eq-mchon-def}
m_{\mathrm{chon}} = \frac{4\pi}{3}\xi_{\mathrm{chon}} a_{\mathrm{chon}}^3
\end{equation}
The number density of chondrules is then
\begin{equation}\label{eq-nchon-afo-t}
n_{\mathrm{chon}}(t) = \frac{\rho_{\mathrm{cloud}}(t)}{m_{\mathrm{chon}}}
\end{equation}
The geometric cross section of a chondrule is $\pi a_{\mathrm{chon}}^2$.
Let us assume that the chondrule has zero albedo. Then the absorption
cross section equals the geometric cross section. The opacity (i.e.
the cross section per gram) is then
\begin{equation}\label{eq-kappa-geometric}
\kappa = \frac{\pi a^2_{\mathrm{chon}}}{
4\pi \xi_{\mathrm{chon}} a_{\mathrm{chon}}^3/3} 
= \frac{3}{4}\frac{1}{\xi_{\mathrm{chon}}a_{\mathrm{chon}}}
\end{equation}
We will assume that the opacity of any possible vapor between the chondrules
will be low compared to that of the chondrules so that it can be
ignored. The optical depth from the center of the cloud to the edge is 
\begin{equation}\label{eq-tau-t}
\tau(t) = \rho_{\mathrm{cloud}}(t) \kappa R_{\mathrm{cloud}}(t)
=\frac{3}{4\pi}\frac{M\kappa}{v_{\mathrm{exp}}^2}\frac{1}{t^2}
\end{equation}
The rate of cooling will strongly depend on this optical depth.

In this paper we will present our results based on a set of fiducial models
as well as parameter scans. The parameters of the fiducial models are listed
in Table \ref{tab-model-param}.

\begin{table}
\begin{center}
\begin{tabular}{|cccc|cc|}
\hline
\hline
Model & $R_{\mathrm{melt,0}}$ & $v_{\mathrm{exp}}$ & $T_0$ & $t_{\mathrm{cool}}$ & $\tau_{\mathrm{cool}}$\\
\hline
F1         & 1 km            & 100 m/s  & 2000 K   & 27 min & 92\\
F2         & 0.1 km          & 1000 m/s  & 2000 K  & 16 sec & 9.2\\
F3         & 10 km           & 100 m/s  & 2000 K   & 7 h & 366 \\
F4         & 0.01 km         & 1000 m/s  & 2000 K  & 1 s & 2.3 \\
\hline
\hline
\end{tabular}
\end{center}
\caption{\label{tab-model-param}The model parameters (first three columns) for the main
  (``fiducial'') models presented in this paper. The cloud mass $M_{\mathrm{cloud}}$
  can be derived from $M_{\mathrm{cloud}}=\tfrac{4\pi}{3}\xi_{\mathrm{chon}}R_{\mathrm{melt,0}}^3$ 
  (cf.\ Eq.~\ref{eq-def-rmelt0}). The last two columns are computed from
    these parameters: the cooling time $t_{\mathrm{cool}}$ (the time after the impact 
    when the cloud starts to cool, see Eq.~\ref{eq-tcool}) and the optical depth at that
    time $\tau_{\mathrm{cool}}\equiv \tau(t=t_{\mathrm{cool}})$ (see Eq.~\ref{eq-taucool}).}
\end{table}

\section{Analytic estimate of the cooling behavior of the chondrule cloud}
\label{sec-estim-cooling}
As the cloud expands it can start to radiate away energy. A fully fledged
time-dependent radiative cooling computation will yield temperature as a
function of time {\em and} space: $T(r,t)$. We will compute this in Section
\ref{sec-radiative-transfer}.

A first estimate of the cooling behavior of the cloud can already be made
with pen and paper. The cloud will initially be optically thick: $\tau\gg
1$. The luminosity of the cloud is then
\begin{equation}
L = 4\pi R_{\mathrm{cloud}}^2\sigma T^4_{\mathrm{eff}}
\end{equation}
where $\sigma$ is the Stefan-Boltzmann constant and $T_{\mathrm{eff}}$ is
the effective surface temperature at $r=R_{\mathrm{cloud}}$. In the above
equation, and from here onward, we omit the $(t)$ for aesthetic reasons, but
the time-dependence is still assumed. During the cooling phase the effective
surface temperature will be lower than the central temperature. A commonly
used estimation of this effect is:
\begin{equation}\label{eq-lum-tau}
L \simeq 4\pi R_{\mathrm{cloud}}^2\sigma T^4\frac{1}{\tau}
\end{equation}
where $T$ is now the central temperature and $1/\tau$ is the correction
factor to account for the lower temperature of the surface
$T_{\mathrm{eff}}$, and is based on radiative diffusion theory which states
that $T\simeq \tau^{1/4}T_{\mathrm{eff}}$ (for $\tau\gg 1$). To compute the
change of the temperature as a result of the radiative loss given by
Eq.~(\ref{eq-lum-tau}) we must use the heat capacity formula. The total
thermal energy stored in the chondrule cloud (approximating it, for the
moment, as an isothermal cloud) is:
\begin{equation}\label{eq-cloud-energy}
E = M_{\mathrm{cloud}} c_m T
\end{equation}
where $c_m$ is the mass-weighted specific heat of the lava droplet. A value
of $c_m=10^7$ erg g$^{-1}$ K$^{-1}$ is a reasonable value which we will
adopt here.  The radiative loss time scale can now be defined as
\begin{equation}\label{eq-trad}
t_{\mathrm{rad}}(t) = \frac{E}{L} = \frac{M_{\mathrm{cloud}} c_m\tau}{4\pi R_{\mathrm{cloud}}^2\sigma T^3}
\end{equation}
This time scale varies with time: it is very long at early times because the
cloud is then still extremely optically thick. At late times the cloud is
optically thinner and the radiation can more freely escape
(Eq.~\ref{eq-lum-tau}) and $t_{\mathrm{rad}}$ becomes smaller. It is
therefore to be expected that at early times the temperature of the
chondrules remains constant. Note that since the cloud consists of liquid
drops that stay at a constant volume and move away from each other, there is
no adiabatic cooling involved here. Any potential vapor may adiabatically
cool, but we will ignore this effect in this paper, assuming that the total
mass in vapor is always small compared to the mass in droplets.

At some point in time, however, $t_{\mathrm{rad}}$ becomes smaller than $t$
and the chondrules start to cool. Based on the results of the true radiative
transfer calculations of Section \ref{sec-radiative-transfer} it turns out
that near the center of the cloud the transition from the initial constant
temperature phase to the temperature-decline phase occurs roughly at a time
$t_{\mathrm{cool}}$ defined by
\begin{equation}\label{eq-tcool-def}
t_{\mathrm{rad}}(t_{\mathrm{cool}}) = 5\,t_{\mathrm{cool}}
\end{equation}
Inserting
Eqs.(\ref{eq-trad},\ref{eq-lum-tau},\ref{eq-cloud-energy},\ref{eq-tau-t})
into Eq.~(\ref{eq-tcool-def}) yields
\begin{equation}\label{eq-tcool}
t_{\mathrm{cool}}=\left(\frac{1}{5}\frac{3}{(4\pi)^2}
\frac{M_{\mathrm{cloud}}^2c_m\kappa}{v_{\mathrm{exp}}^4\sigma
T_0^3}\right)^{1/5}
\end{equation}
Note that in Eq.~(\ref{eq-tcool-def}) the factor of 5 is purely 
  empirical. The estimate we make here is essentially a dimensional 
  analysis in which proportionality factors have to be calibrated
  against more exact calculations (in our case the full radiative
  transfer calculation).

Roughly for $t<t_{\mathrm{cool}}$ the temperature stays constant while
for $t>t_{\mathrm{cool}}$ the temperature drops with time. From now on we
shall define {\em the} $t_{\mathrm{cool}}$ as the one calculated for the
center of the cloud. Near the surface the cooling sets in earlier. 

\begin{figure}
\begin{center}
\includegraphics[width=0.5\textwidth]{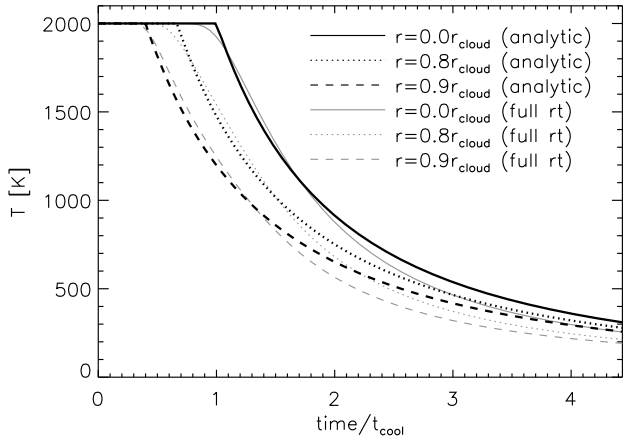}
\end{center}
\caption{\label{fig-temp-afo-time-analytic}The temperature evolution of the
  expanding cloud model according to the analytical estimate of
  Eqs.~(\ref{eq-t-analytic},\ref{eq-t-analytic-zero}), shown in black. In
  grey are, for comparison, the full radiative transfer solutions from
  Section \ref{sec-radiative-transfer} for the fiducial model F1 (see Table
  \ref{tab-model-param}). The temperatures are shown at three positions in
  the cloud: the center, at 80\% of the radius and at 90\% of the radius
  (near the surface of the cloud). The time scale is scaled to
  $t_{\mathrm{cool}}$ (Eq.~\ref{eq-tcool}). The analytic solutions are only
  valid as long as $\tau_{\mathrm{cool}}\gg 1$ (cf.\ Eq.~\ref{eq-taucool}).}
\end{figure}

The temperature decline with time can also be estimated, at least up until
the point where the cloud becomes optically thin, after which our
approximation breaks down. The way to make this estimate is to solve the
cooling equation
\begin{equation}\label{eq-temp-cool-analyt}
\frac{dT(t)}{dt} = -\frac{T(t)}{t_{\mathrm{cool}}(t)}
\end{equation}
where $t_{\mathrm{cool}}(t)$ is the cooling time given by
Eq.~(\ref{eq-tcool}) but with $T_0$ replaced by $T(t)$. The reasoning behind
this is that if we would instead use $t_{\mathrm{rad}}(t)$ (which might look
more reasonable at first sight) we will quickly cool to temperatures for
which condition $t_{\mathrm{rad}}\lesssim t$ is again no longer
fulfilled. This is because $t_{\mathrm{rad}}\propto T^{-3}$ and thus very
rapidly rises with declining temperature.  If we insert Eq.~(\ref{eq-tcool})
with $T_0$ replaced by $T$ into Eq.~(\ref{eq-temp-cool-analyt}) we obtain a
differential equation that can be solved by separation of variables. The
solution is
\begin{equation}\label{eq-t-analytic}
T(t>t_{\mathrm{cool}}) = T_0\left[\frac{3}{5}\frac{t}{t_{\mathrm{cool}}}+
\frac{2}{5}\right]^{-5/3}
\end{equation}
where here $t_{\mathrm{cool}}$ is again the original one with $T_0$. We assume
that for $t<t_{\mathrm{cool}}$ we have:
\begin{equation}\label{eq-t-analytic-zero}
T(t<t_{\mathrm{cool}}) = T_0
\end{equation}
The solution Eqs.~(\ref{eq-t-analytic},\ref{eq-t-analytic-zero}) holds true
for the central temperature. But most of the mass of a homogeneous sphere
resides in the outer parts. For the temperature at $r=0.8R_{\mathrm{cloud}}$
and $r=0.9R_{\mathrm{cloud}}$ (roughly the radii of half mass and of 75\%
mass respectively) the best fitting solution to the real solutions of
Section \ref{sec-radiative-transfer} are similar to
Eq.~(\ref{eq-t-analytic}) but with the term $2/5$ replaced by $3/5$ and
$3.8/5$ respectively. These analytic estimates of the temperature as a
function of time are plotted in Fig.~\ref{fig-temp-afo-time-analytic}. 
  As one can see, although the time after the impact at which the cooling
  stars is different depending on whether you look at the center or at the
  edge of the cloud, the typical cooling rate after the cooling begins is
  similar throughout the cloud.

It is important to remind ourselves that the cooling normally starts
  well before the cloud becomes optically thin. Optically thin cooling would
  be extremely fast: a single chondrule would cool within about 1 second.
  The protracted cooling over a time scale of hours occurs because the high
  optical depth (even after $t=t_{\mathrm{cool}}$) acts as a kind of blanket
  that keeps the chondrules warm, albeit a blanket that is thinning over
  time as the optical depth drops. To get a feeling for this we can
  calculate the optical depth at the time $t=t_{\mathrm{cool}}$ from
Eq.~(\ref{eq-tau-t}):
\begin{equation}\label{eq-taucool}
\tau_{\mathrm{cool}}\equiv \tau(t_{\mathrm{cool}}) =
\frac{3^{3/5}5^{2/5}M_{\mathrm{cloud}}^{1/5}\kappa^{3/5}\sigma^{2/5}T_0^{6/5}}
{(4\pi)^{1/5}v_{\mathrm{exp}}^{2/5}c_m^{2/5}}
\end{equation}
So Eqs.~(\ref{eq-t-analytic},\ref{eq-t-analytic-zero}), which are based
  on the assumption of the ``blanket effect'' of high optical depth, are
expected to be valid for the case $\tau_{\mathrm{cool}}\gg 1$. In table
  \ref{tab-model-param} one can see that $\tau_{\mathrm{cool}}$ is well
  above 1 for all models, except the most extreme model F4 which has anyway
  a much too small cooling time to be consistent with chondrule textures.

\begin{figure}
\begin{center}
\includegraphics[width=0.5\textwidth]{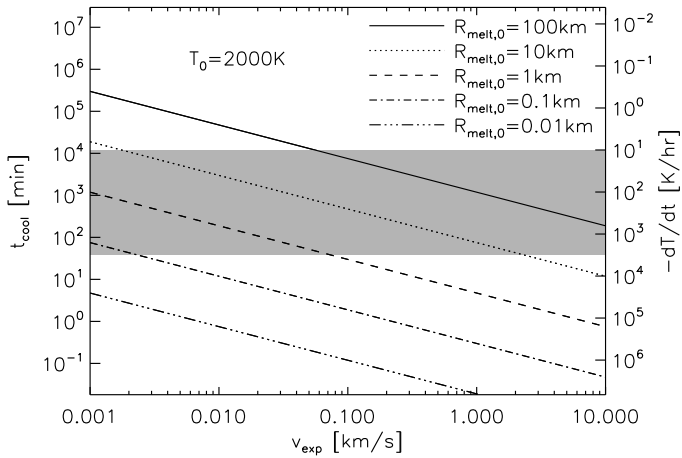}
\end{center}
\caption{\label{fig-tcool-afo-params}The cooling time $t_{\mathrm{cool}}$
  (Eq.~\ref{eq-tcool}) as a function of parameters of the model for
  $T_0=2000$ K and $a_{\mathrm{chon}}=0.03$ cm. Note that the time to
    cool down from 2000 K to 1400 K ($t_{1400}$) is, according to
    Eq.~(\ref{eq-t1400-definition}), about four times smaller than
    $t_{\mathrm{cool}}$. On the right axis the corresponding cooling rate
  $|dT/dt|$ in K/hr is plotted. The grey area is the observed range of
  chondrule cooling rates (see main text). Note that the values of
  $t_{\mathrm{cool}}$ are (by definition) for the center of the cloud. The
  regions of the cloud more closely to the surface will start to cool
  earlier, but the cooling rate is almost the same.}
\end{figure}

The analytic model of Eqs.~(\ref{eq-t-analytic},\ref{eq-t-analytic-zero})
tells us that for $t>t_{\mathrm{cool}}$ the temperature drops off suddenly
on a typical time scale of $t_{\mathrm{cool}}$. The temperature decay rate
at $t=t_{\mathrm{cool}}$ is:
\begin{equation}
\dot T(t=t_{\mathrm{cool}})\equiv\left.\frac{dT}{dt}\right|_{t=t_{\mathrm{cool}}} = -\frac{T_0}{t_{\mathrm{cool}}}
\end{equation}
This is plotted, together with $t_{\mathrm{cool}}$, 
  in Fig.~\ref{fig-tcool-afo-params}. It is also 
  useful to define another kind of cooling time scale $t_{1400}$ which 
  is the time between the start of the cooling at $T=T_0$ and the time
  when the temperature has dropped below the solidus temperature 1400 K. 
  To first order we can write
\begin{equation}\label{eq-t1400-definition}
t_{1400} \simeq \frac{T_0-1400K}{|\dot T(t=t_{\mathrm{cool}})|}=\left(1-\frac{1400K}{T_0}\right)t_{\mathrm{cool}}
\end{equation}
The quantities $\dot T$ and $t_{1400}$ can be compared to the constraints
coming from the analysis of textures of chondrules, which put time limits on
the cooling process of order of hours, or more precisely: cooling rates in
the range $10\cdots \sim 3000$ K/hr (see Morris \& Desch 2010 and 
Desch et al. 2012 and references
therein, though see also Miura \& Yamamoto 2014 for a different view
based on theoretical modeling of crystallization), indicated by the grey area in
Fig.~\ref{fig-tcool-afo-params}. According to Eq.~(\ref{eq-tcool})
$t_{\mathrm{cool}}$ goes as $M_{\mathrm{cloud}}^{2/5}$ and as
$v_{\mathrm{exp}}^{-4/5}$. There is not very much freedom of choice of
$T_0$: it must lie somewhere between $1770$ and $2120$ K (see Morris \&
Desch 2010 and references therein). The $\kappa$ does not have too much
wiggle room either: the radii of chondrules are known. This means that
constraints on $t_{\mathrm{cool}}$ from textural analysis directly set
limits on the ratio $M_{\mathrm{cloud}}/v_{\mathrm{exp}}^2$ or equivalently
on the ratio $R_{\mathrm{melt,0}}^3/v_{\mathrm{exp}}^2$. If, for example, we
choose $a_{\mathrm{chon}}=0.03$ cm and $T_0=2000$ K we obtain $\kappa=7.6$
cm$^2$/g. If we require, for instance, the cooling time scale to be
$t_{\mathrm{cool}}=10$ minutes, then we find that
$M_{\mathrm{cloud}}/v_{\mathrm{exp}}^2\simeq 10^7$ g s$^2$ cm$^{-2}$. This
would be fulfilled e.g.\ for $v_{\mathrm{exp}}=1$ km/s and
$M_{\mathrm{cloud}}=10^{17}$ g (a mass corresponding to a ball of magma of
roughly $R_{\mathrm{melt,0}}=$2 km radius). It would also be fulfilled by a
smaller mass and smaller velocity: e.g.\ for $v_{\mathrm{exp}}=10$ m/s and
$M_{\mathrm{cloud}}=10^{13}$ g ($R_{\mathrm{melt,0}}=$0.1 km radius). Such
small mass could either mean that upon impact only a small fraction of the
debris is in the form of chondrule droplets, or it could simply mean that
the largest closed unit of the droplet splash (subcloud) is so small, but
many such expanding cloudlets of chondrules are ejected. The elongated
streams of debris found in Asphaug et al.~(2011) could be regarded as a
string of such smaller mass cloudlets.

\begin{figure}
\begin{center}
\includegraphics[width=0.5\textwidth]{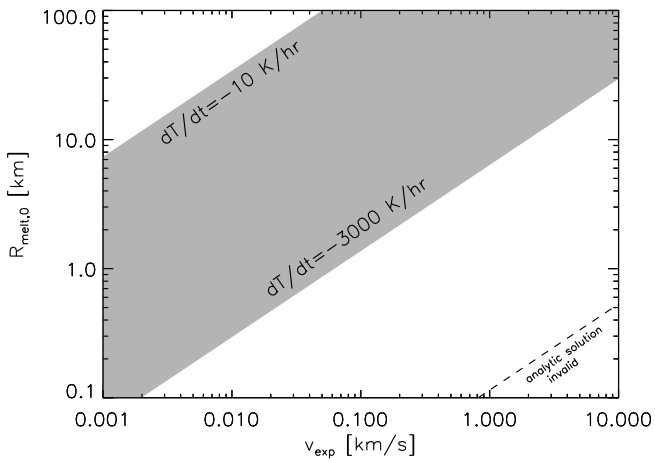}
\end{center}
\caption{\label{fig-rmelt-afo-vexp}The parameter space of the model.
The grey area shows where the solutions have cooling rates in the
range observed for chondrules. In the bottom-right the region is
shown where $\tau_{\mathrm{cool}}<10$, where the analytic solution
becomes invalid.}
\end{figure}

This is shown more directly in Fig.\ \ref{fig-rmelt-afo-vexp} which 
shows the parameter space of the model, with the grey area again showing
the models which give cooling rates between 10 and 3000 K/hr. 

\section{Time-dependent radiative transfer}
\label{sec-radiative-transfer}
So far we have only made an estimate of the cooling behavior of the cloud of
liquid chondrules. Let us now calculate the temperature profile with a full
treatment of time-dependent radiative transfer. The time-dependence comes in
only due to the time it takes the lava droplets to convert their heat into
radiation, not in the form of light-travel time (which typically only plays
a role for media at temperatures above $10^5$K). This can be easily verified
by comparing the radiative energy density $aT^4$ to the material energy
density $\rho c_m T$, which in our case always satisfies $aT^4\ll \rho c_m
T$. Therefore the radiative transfer equation itself is stationary, and
the time-dependence is in the equation for heating/cooling of the chondrules.

\subsection{Equations of time-dependent radiative transfer} 
\label{sec-rt-eqs}
The formal radiative transfer equation is (see e.g.\ Mihalas \& Mihalas
1999):
\begin{equation}\label{eq-formal-rt-eq}
\vec n\cdot \vec\nabla I(\vec x,\vec n)=\rho\kappa (B(\vec x)-I(\vec x,\vec n))
\end{equation}
where $I$ is the frequency-integrated mean intensity in erg s$^{-1}$
cm$^{-2}$ ster$^{-1}$, $\vec x$ is the position vector in space, $\vec n$ is
the unit vector of direction of the radiation, and finally $B$ the
frequency-integrated Planck function given by
\begin{equation}\label{eq-freqint-planck}
B(T) = \frac{\sigma}{\pi}T^4
\end{equation}
We will use frequency-integrated quantities because the opacity of a
chondrule is expected to be roughly equal to the geometric cross section,
independent of frequency, because the chondrules are much larger than the
wavelength of infrared radiation. In this case the frequency-dependence of
the radiative transfer does not need to be
considered. Eq.~(\ref{eq-formal-rt-eq}) is called the ``formal transfer
equation'', and describes the transport of radiation along rays of direction
$\vec n$. It is easy to integrate {\em if} the value of $B(\vec x)=B(T(\vec
x))$ is known everywhere, but since we want to calculate the temperature
$T(\vec x)$, this function is part of the thing we want to solve.

In our 1-D spherically symmetric setting Eq.~(\ref{eq-formal-rt-eq})
can be written as
\begin{equation}\label{eq-formal-rt-eq-spher}
\mu\frac{dI_\mu(r)}{dr}+\frac{1-\mu^2}{r}\frac{dI_\mu(r)}{d\mu}=\alpha (B(T(r))-I_\mu(r))
\end{equation}
where we shortened
\begin{equation}\label{eq-def-alpha-opac}
\rho\kappa =: \alpha
\end{equation}
Here $\mu=\cos(\theta)$ where $\theta$ is the angle between the ray along
which the radiation is followed and the radially outward pointing unit
vector. The Eddington factor method of solving this equation relies on the
definition of the first three angular moments of the intensity at each
location $r$:
\begin{eqnarray}
J(r) &=& \frac{1}{2}\int_{-1}^{+1}I_\mu(r)d\mu\label{eq-radmom-j}\\
H(r) &=& \frac{1}{2}\int_{-1}^{+1}I_\mu(r)\mu d\mu\label{eq-radmom-h}\\
K(r) &=& \frac{1}{2}\int_{-1}^{+1}I_\mu(r)\mu^2 d\mu\label{eq-radmom-k}
\end{eqnarray}
Here $J$ is called the mean intensity and $H$ is the flux divided by $4\pi$.
By integrating Eq.~(\ref{eq-formal-rt-eq-spher}) over $\tfrac{1}{2}d\mu$
after multiplying it by $1$ and by $\mu$ respectively one obtains the first
two moment equations:
\begin{eqnarray}
\frac{1}{r^2}\frac{d(r^2H)}{dr} &=& \alpha(B(T)-J)\\
\frac{dK}{dr}+\frac{3f-1}{r}J &=& -\alpha H
\end{eqnarray}
These are two equations with three unknowns. To close this set of equations
we write
\begin{equation}\label{eq-k-afo-j-eddington}
K(r) = f(r)J(r)
\end{equation}
where $f$ is called the ``Eddington factor''. If $f(r)$ is known, then the
two moment equations are in closed form and can be solved for $J(r)$ and
$H(r)$. From the moment equations alone we cannot know what $f(r)$ is for
all $r$, but for certain limiting cases we know their values: For optically
thick media $f=1/3$. Outside of the cloud for $r\rightarrow\infty$ we will
get $f\rightarrow 1$, which is the free-streaming limit. However, for
general $r$ and for general values of the optical depth we must employ
another method of calculating $f(r)$. We do this by integrating, for the
current value of $T(r)$ (and thus $B(T(r))$), the formal transfer equation
Eq.~(\ref{eq-formal-rt-eq-spher}) using the ``tangent ray method'' (see
e.g.~Mihalas \& Mihalas 1999). By integrating the resulting
intensities over angle we calculate the current estimates of $J(r)$ and
$K(r)$. Let us call these $J_{\mathrm{fte}}(r)$ and $K_{\mathrm{fte}}(r)$
(where ``fte'' stands for ``formal transfer equation''). Our current 
estimate of $f(r)$ is then
\begin{equation}\label{eq-eddfact-kj}
f(r) =\frac{K_{\mathrm{fte}}(r)}{J_{\mathrm{fte}}(r)}
\end{equation}
This is then the $f(r)$ function we stick into
Eq.~(\ref{eq-k-afo-j-eddington}), so that the moment equations can be
solved for $J(r)$ and $H(r)$. We also need to impose boundary conditions
at the inner and outer edge. At the inner edge we take $H=0$ (zero flux).
At the outer edge we could set $H=J$, which is valid if the outer edge of
our computational domain is at $r\rightarrow\infty$. For finite outer
radius we set instead
\begin{equation}\label{eq-outer-bc}
H(r_{\mathrm{out}}) = \frac{H_{\mathrm{fte}}(r_{\mathrm{out}})}{J_{\mathrm{fte}}(r_{\mathrm{out}})}J(r_{\mathrm{out}})
\end{equation}
Note that we can choose to set the outer edge of our computational
domain $r_{\mathrm{out}}$ at $r_{\mathrm{out}}=R_{\mathrm{cloud}}$, but we
can also set it at $r_{\mathrm{out}}>R_{\mathrm{cloud}}$, as long as we 
properly set the boundary condition at $r=r_{\mathrm{out}}$ according to
Eq.~(\ref{eq-outer-bc}).

We can now combine the two moment equations into the following form:
\begin{equation}\label{eq-secorder-momeq}
\frac{1}{\alpha}\frac{1}{r^2}\frac{d}{dr}\left\{
\frac{r^2}{\alpha}\left[\frac{d(fJ)}{dr}+\frac{3f-1}{r}J\right]
\right\}=J-\frac{\sigma}{\pi}T^4
\end{equation}
This can be solved, together with the boundary conditions, using a matrix
equation. The details of this are discussed in Appendix
\ref{app-rt-eq-discrete}.

Next we must compute how the temperature $T(r)$ reacts to the radiation
field, or in other words, how $T(r)$ radiatively cools with time. 
The energy equation is
\begin{equation}\label{eq-dtemp-dt}
\rho c_m\frac{dT}{dt}=4\pi\alpha\left(J-\frac{\sigma}{\pi}T^4\right)
\end{equation}
where $\rho$ is given by Eq.~(\ref{eq-rho-afo-time}). The $d/dt$ operator is
the comoving derivative, i.e.\ the time derivative is computed for a given
chondrule. Eq.~(\ref{eq-dtemp-dt}) is a local equation at each location. The
non-locality of the heating/cooling is only through the non-locality of the
equation for $J(r)$ (Eq.~\ref{eq-secorder-momeq}).

The way we solve these equations numerically is by setting up a radial grid
of $N$ gridpoints $r_i$ (with $1\le i\le N$) upon which we define the mean
intensity $J_i$ and the temperature $T_i$. Since the cloud is expanding we
let the gridpoints move along with the material (Lagrangian approach) so
that the location of grid point $i$ at time $t$ is
\begin{equation}
r_i(t) = \eta_iv_{\mathrm{exp}}t
\end{equation}
where $\eta$ is the dimensionless radial coordinate
\begin{equation}\label{eq-eta-def}
\eta = \frac{r}{R_{\mathrm{cloud}}(t)} = \frac{r}{v_{\mathrm{exp}}t}
\end{equation}
The values of $\eta_i$ do not change with time. Each chondrule stays at
constant $\eta$, and so the comoving time derivative $d/dt$ used in e.g.
Eq.~(\ref{eq-dtemp-dt}) becomes simply the time derivative of that quantity
at a given $\eta_i$ grid point.

In principle one could imagine a simple time-dependent numerical integration
method for the combination of
Eqs.(\ref{eq-secorder-momeq},\ref{eq-dtemp-dt}). We know $J_i^n$ and $T_i^n$
at time $t=t_n$, and we can solve for $T_i^{n+1}$ by taking an explicit
Euler integration step of Eq.~(\ref{eq-dtemp-dt}). Then we can solve the
moment equations (because now we have $B_i^{n+1}=B(T_i^{n+1})$) and obtain
$J_i^{n+1}$. The problem with this scheme is that the required time steps
can become very small due to numerical stiffness. A more robust method is to
integrate the complete system
Eqs.(\ref{eq-secorder-momeq},\ref{eq-dtemp-dt}) using an implicit
integration scheme. This is discussed in appendix \ref{app-rt-eq-discrete}.

\subsection{Results of the time-dependent radiative transfer models}
\label{sec-rt-results}
\begin{figure*}
\begin{center}
\includegraphics[width=0.47\textwidth]{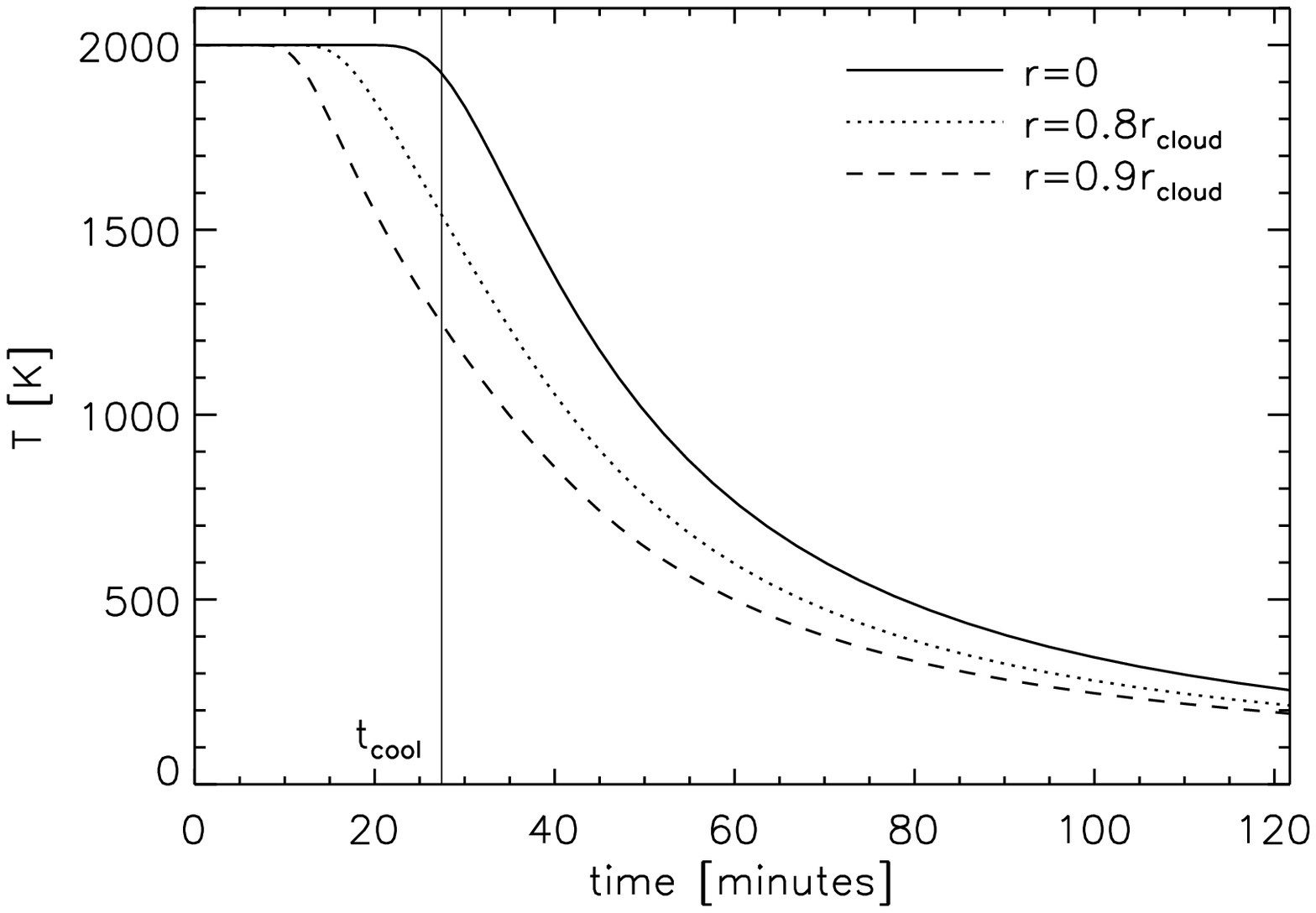}
\includegraphics[width=0.47\textwidth]{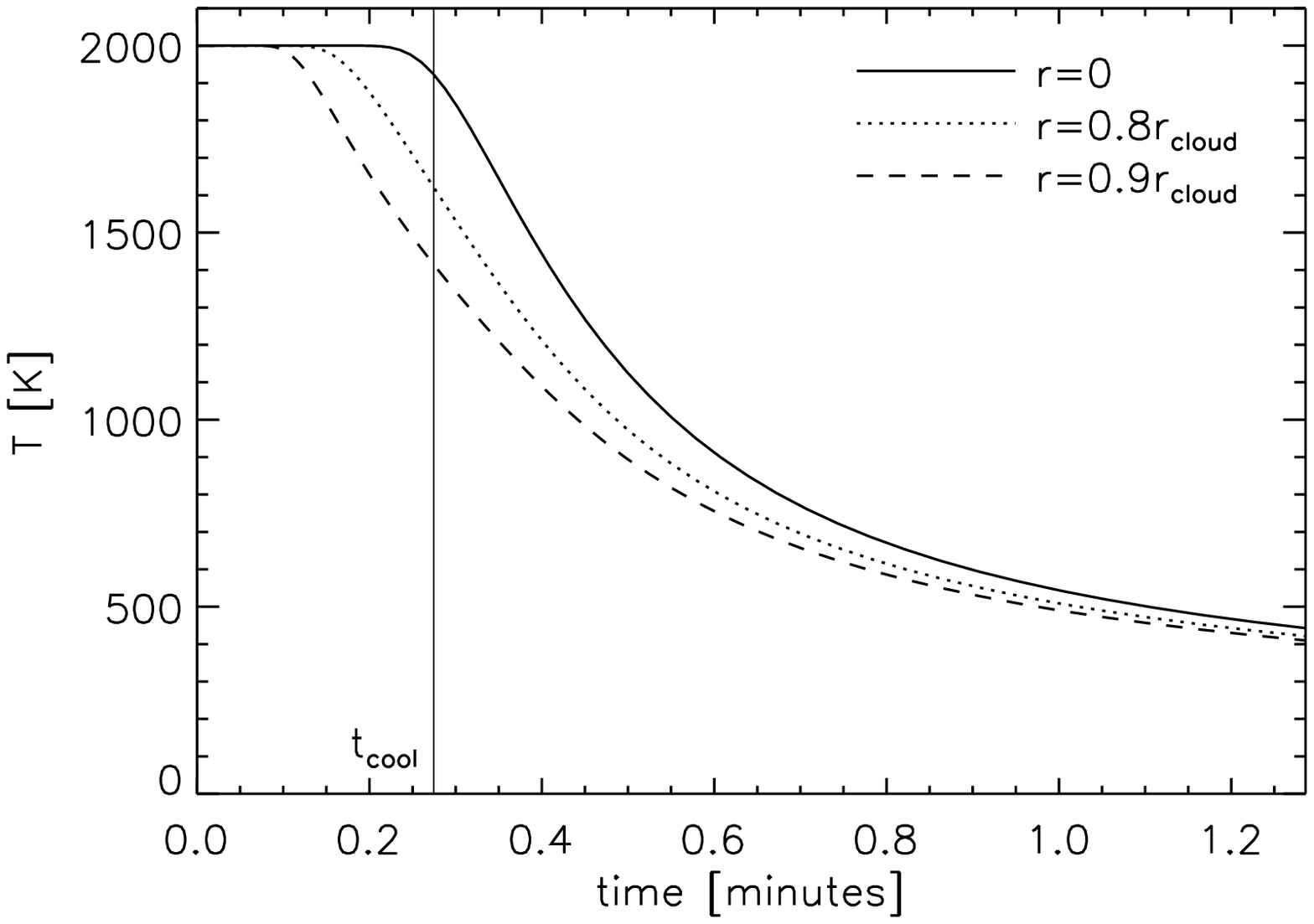}\\
\includegraphics[width=0.47\textwidth]{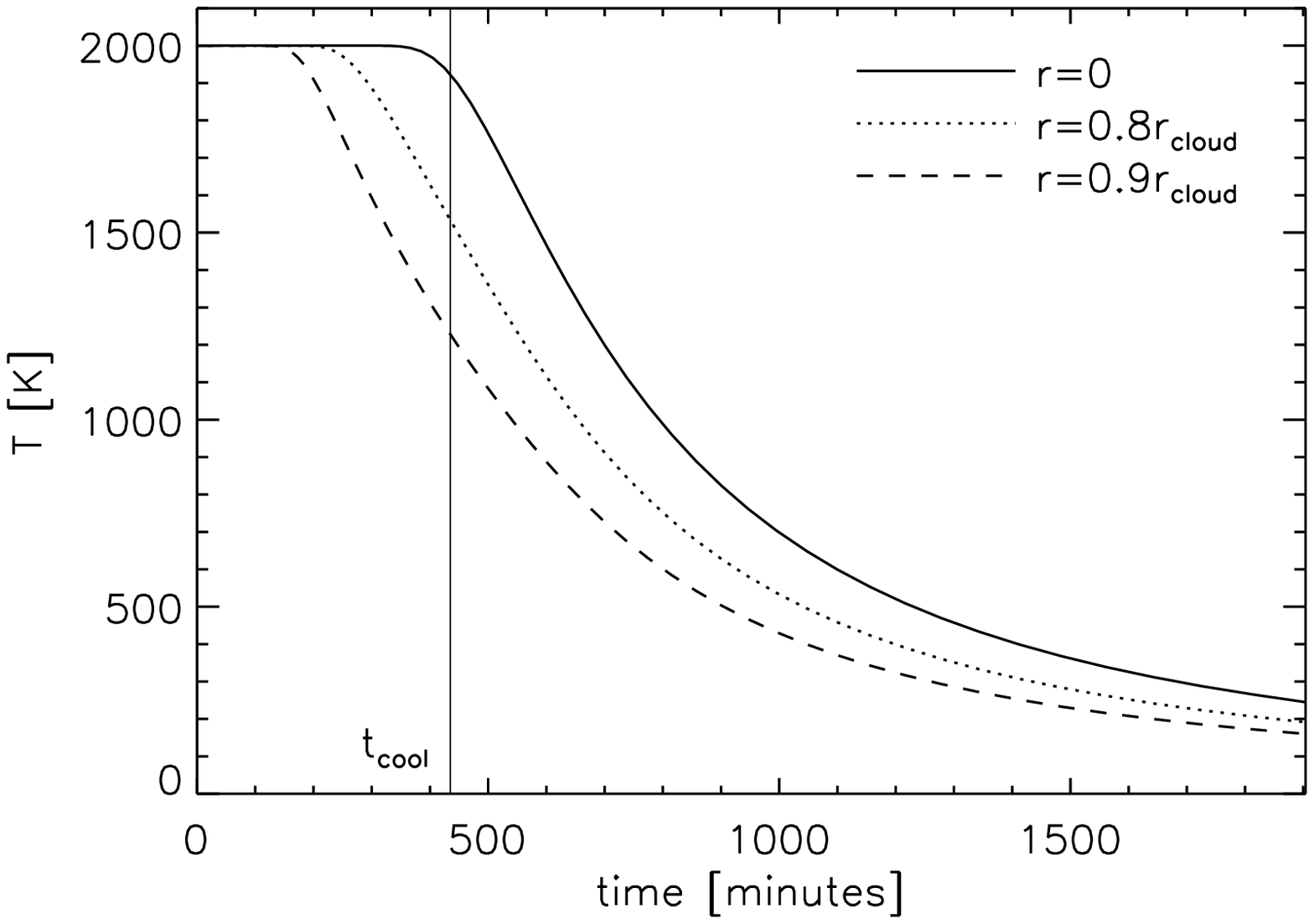}
\includegraphics[width=0.47\textwidth]{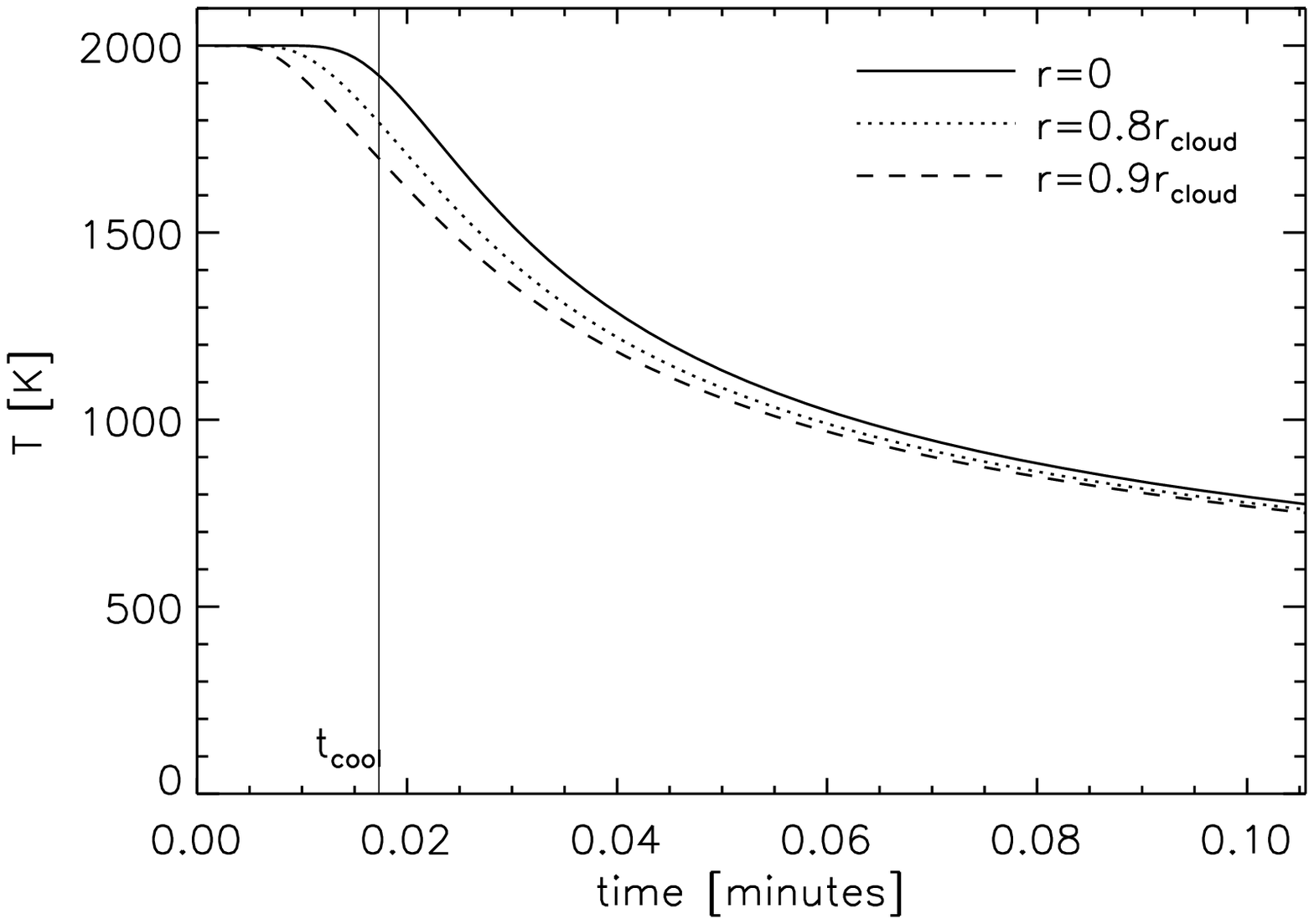}
\end{center}
\caption{\label{fig-tprofile-1}The temperature evolution of the fiducial
  expanding cloud models F1 (top-left), F2 (top-right), F3 (bottom-left) and
  F4 (bottom-right). See Table \ref{tab-model-param} for the parameters. The
  temperatures are shown at three positions in the cloud: the center, at
  80\% of the radius and at 90\% of the radius (near the surface of the
  cloud). The vertical line shows the time $t_{\mathrm{cool}}$ of
  Eq.~(\ref{eq-tcool}).}
\end{figure*}

The results of the time-dependent radiative transfer calculations are shown
in Fig.~\ref{fig-tprofile-1}.
These results confirm the estimates made in Section \ref{sec-estim-cooling}.
We find that the shape of the cooling curves is mostly the same for
all models, except for the time scaling, which we know from the analytically
expression of $t_{\mathrm{cool}}$ (Eq.~\ref{eq-tcool}). Only for the rather
extreme (and presumably unrealistic) case such as fiducial model F4 with
$R_{\mathrm{melt,0}}=0.01$ km and $v_{\mathrm{exp}}=1000$ m/s we find from
the full radiative transfer model that the cooling is slower than the
analytic solution. This is because in that case the condition that
$\tau_{\mathrm{cool}}$ (Eq.~\ref{eq-taucool}) is $\gg 1$ is broken and the
cooling time scale will then not be determined by the optical depth decline
but by the effectiveness by which a chondrule can convert its heat into
radiation. These cases, however, imply typically time scales of seconds
rather than hours, and these are anyway too short to be
consistent with chondrule textures. 

It seems that the analytic solutions are fairly accurate when
$\tau_{\mathrm{cool}}\gtrsim 10$, which is always true for the interesting
regime of parameter space.

From Figs.~\ref{fig-temp-afo-time-analytic} and \ref{fig-tprofile-1} we see
the model systematically predicts a period of constant temperature with a
duration similar to the later cooling phase. This is true at least near the
center of the cloud. Near the surface this temperature plateau is shorter,
but the cooling time is similar. 

\begin{figure}
\begin{center}
\includegraphics[width=0.5\textwidth]{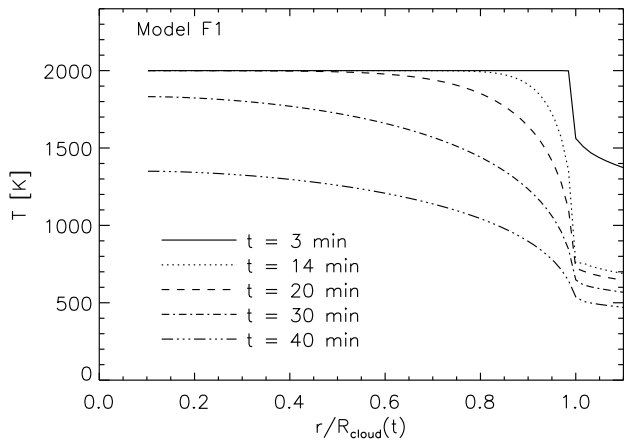}
\end{center}
\caption{\label{fig-t-afo-r}The radial temperature temperature structure
of fiducial model F1 for different times. The horizontal axis is the 
radius relative to the outer cloud radius. Note that the temperature 
profile extends beyond the outer radius of the cloud because even if
that region is empty, a chondrule residing there would have a 
well-defined temperature. For our purposes, however, only the 
temperature within the cloud is relevant.}
\end{figure}

In Fig.\ \ref{fig-t-afo-r} we show, for model F1, how the temperature
  looks as a function of radial position in the cloud, for different times
  after the collision. Since the cloud is expanding, which would make it
  difficult to compare the temperature profiles at different times, we use
  the radial coordinate relative to the outer cloud radius. One sees that
  the outer regions start cooling earlier because the optical depth to the
  surface is smaller. Also one sees that a negative radial temperature
  gradient is produced which causes the radiative diffusive energy transport
  to the surface.

The cooling curve from our model is significantly different from the
ones predicted from shock heating (e.g.\ Morris \& Desch 2010, Morris et
al.\ 2012). The shock heating models predict a radiative preheating phase, a
temperature spike with rapid cooling, and depending on geometric conditions
followed by a slower cooling phase. In our model the temperature is high
from the start (or for high-speed collisions: upon impact), stays relatively
constant for some time and then drops over a similar time scale.

\section{Discussion}
The scenario of splashes of melt droplets originating from colliding
pre-molten planetesimals has been discussed at length in several papers by
Sanders and coworkers (e.g.\ Sanders \& Scott 2012) and the paper by
  Asphaug et al.~(2011). These papers discuss how this scenario holds up
agains the many meteoritic constraints.  We will not repeat these arguments
here, but refer for those discussions to those papers. Instead, we will
focus on the discussion of aspects related to our splash cooling model.

\subsection{Splash geometry} 
Our model describes a simple spherical expanding cloud as a model for
  (part of) the splash resulting from colliding bodies. It is obvious that
  this is a drastic simplification. The SPH model of Asphaug et al.~(2011)
  shows a much more realistic geometry involving, among other complexities,
  tube-like geometries. A proper 3-D time-dependent radiative transfer
  simulation done in concert with the gravito-hydrodynamic simulation of the
  impact splashing is required to know precisely what the cooling behavior
  is. Nevertheless we predict that our result can be used to estimate the
  results of those detailed computations: the tube-like geometry can be
  approximated as a homologously expanding cylinder, expanding not only in
  the 2 perpendicular directions but also in the longitudinal direction. Its
  properties are then not too much different from a chain of spherically
  expanding clouds with diameters similar to the diameter of the tube,
  except that these spherical cloud cannot cool in all directions ($4\pi$)
  but only in part of the sky (in the other part it will heat its neighbors
  and be heated by its neighbors). We would then expect that the cooling
  time $t_{\mathrm{cool}}$ will be a bit longer but not by a large factor.

Also, one can expect some parts of the splash to be ejected at slow
  speed compared to the escape speed, so that it may fall back to the
  remaining (unsplashed) bodies and accrete on them. This evidently breaks
  the assumption of homologous expansion, and as a result will also deform
  any initially spherical sub-cloud into a more pancake-like shape. The idea
  of regarding the splash cloud as a collection of spherical sub-clouds is
  then clearly no longer valid, not even in an approximative way. In such
  cases a full 3-D radiation-gravito-hydrodynamic simulation of the splash
  is unavoidable. However, multi-dimensional radiation-hydrodynamics is an
  enormously challenging problem, in particular for problems involving such
  extremely complex geometries. It is an interesting challenge for the
  future. For getting order-of-magnitude estimates we may be forced to
  stick, for the time being, with simple models such as the one presented in
  this paper.

\subsection{Chondrule sizes}
According to the model by Asphaug et al.~(2011), by which the lava drop size
is calculated from the initial pressure under which the magma was in the
pre-molten planetesimals before the collision, the planetesimal sizes must
have been 10 km or larger. Since our cloud of lava droplets can be one of a
number of such clouds ejected from the impact, the $R_{\mathrm{melt,0}}$ of
our model cannot be directly compared to that $>$ 10 km size lower limit
from Asphaug. But it does give an indication that, if we adopt their model
of the chondrule sizes, it seems reasonable to be looking to values of
  roughly the order of $R_{\mathrm{melt,0}}\simeq 10$ km or
  larger. To elucidate our reasoning: If we collide two fully molten
  planetesimals of 20 km radius, and we model the splash with 10 spherically
  expanding droplet clouds, then we would get $R_{\mathrm{melt,0}}\simeq 12$
  km. In Fig.~\ref{fig-tcool-afo-params} this means we are looking at the
region of the diagram above the dotted line. 

There is a caveat, however: What tells us that it is just 10 clouds,
  not a 1000? This brings us to the complicated issue of the actual geometry
  of the splash. If we ``decompose'' the splash geometry into many small
  spherical clouds we must assure that these clouds can radiatively cool to
  the outside. In other words: we cannot consider a big spherical cloud as
  consisting of many smaller sub-clouds, because they will irradiate each
  other and not cool much. But if the cloud geometry is a long extended
  streak, it can be considered as consisting of a chain of spherical
  subclouds. We are here confronted with the limitation of our spherical
  expanding cloud model.

\subsection{Partial pre-melting}
The reason why pre-melting seems to be necessary for the impact splash
scenario to work is that very high speed collisions ($\gtrsim 5$ km/s)
needed to produce melt from cold planetesimals are presumably rare.
However, pre-melting may lead to differentiation of the planetesimals,
causing the iron and nickel to sink into the core and producing a basaltic
mantle. Chondrules formed from such objects would be very non-solar in
composition. Although Sanders \& Scott (2012) argue that convection
reduces the efficiency of the differentiation and that some amount of
differentiation is in fact consistent with e.g.\ L and LL chondrites, 
it seems that this issue is not yet conclusively solved and
more detailed and quantitative modeling of the differentiation process
in these molten planetesimals may need to be done.

However, a middle way is if planetesimals were pre-heated to elevated
temperatures below the melting point. This is to be expected in
  particular for later generation planetesimals (i.e.\ planetesimals formed
  at a time when a substantial fraction of the $^{26}$Al has already
  decayed). Lower velocity collisions might then be sufficient to add the
remaining energy needed to produce melt. The problem with this scenario is
that it requires fine-tuning: The planetesimal must have formed at a time
when the remaining abundance of $^{26}$Al is low enough not to melt the
planetesimal but high enough to heat it to temperatures only slightly below
the melting temperature. This might be too much coincidence.

On the other hand, we know that some planetesimals were strongly
differentiated: the parent bodies of iron and basaltic meteorites. If
chondrules formed from collisions between planetesimals, one would expect
some chondrules to have been formed from collisions with such differentiated
planetesimals. Some chondrules should have compositions that are very iron
poor, presumably more iron-poor than LL-chondrites. This raises the
question where are these? One possible answer is that chondrules produced
in these very early phases ($<$ 1 million years after CAI formation) were
either accreted into the sun or reincorporated into other bodies, where
they were again molten. But if this process of elimination is so efficient,
why did some CAIs survive and get incorporated into the (later) chodrites?
This shows that while the impact splashing model is appealing, there is
more work to be done to answer such questions.

\subsection{Compound chondrules}
\label{sec-comp-chond}
In our homologously expanding cloud model the melt droplets move away from
their nearest neighbor at extremely low velocity (millimeters per minute).
Small inhomogeneities (which are of course unavoidable in such a messy
process as an impact) would lead to some droplets actually colliding. 
When they do so during the initial constant temperature phase, they 
presumably coalesce and form a larger melt droplet. However, when they 
collide during the cooling phase, they may have become cool enough to be
plastic but not liquid anymore. They will then produce a compound chondrule.

It would be useful if we could quantify the percentage of chondrules
  that become compound in this way. However, at present we see no way how we
  can estimate the random relative velocities between adjacent chondrules in
  the expanding clouds.

Many compound chondrules, however, clearly show one chondrule having
remained rigid, while the other having been indented. So clearly they were
of different temperature. This is difficult to achieve in our simple
spherical expanding cloud model. For this it will be necessary to have other
debris (e.g.\ another cloud-fragment emerging from the impact site) to
interdisperse with our cloud of droplet, so as to have 
chondrules from different thermal environments to mix. And this must 
happen on a time scale shorter than the time scale of thermal equilibration
between the chondrules from these different origins. To make this
more quantitative would be, however, a challenge. It would likely
require us to contemplate complex splash geometries.

Metzler (2012) found that some chondrites in fact contain entire large scale
clusters of compound chondrules. A possible explanation could be that some
of the impact debris may have fallen back onto the remainder of the biggest
of the two colliding planetesimals. If this happens quickly enough, so that
this re-accretion occurs before the chondrules have cooled below the
temperature at which the chondrules are still sticky, then it is reasonable
to assume that such clusters may form as chondrules simply fall onto each
other.

\subsection{What about relict grains and rims?}
\label{sec-relict-grains}
Another problem with our simple spherically symmetric expanding plume model
is that it does not account for any possible ``pollution'' of the melt
droplets with other (non-molten) debris. Some chondrules contain relict
grains (Jones 1996), which must have been inserted while the chondrule was
still liquid, but the chondrule must have cooled very soon thereafer. As in
the case with the compound chondrules (Section \ref{sec-comp-chond}) this
would require debris from a different part of the impact debris cloud to
interdisperse with our cloud of droplets. This foreign debris must have
originated from non-molten parts of the original planetesimals, e.g.\ the
regolith or pristine dust covering the crust of the original planetesimals.
Only those grains from that non-molten debris that entered late enough can
enter the chondrules shortly before they cool below the solidus, and thus
survive. Like with the compound chondrules, however, getting quantitative
estimates of this process is challenging.

\subsection{Producing chondritic parent bodies}
The impact model may provide a plausible scenario for the {\em formation} of
chondrules, but how do these chondrules (and matrix) accrete to form another
planetesimal (the parent body of the later chondritic meteorite)? A nice
aspect of the low-velocity collision of pre-molten planetesimals model is
that much of the debris may, in fact, reaccrete onto one of the surviving
original planetesimals (or what remains of it). The idea is that due to the
high gas density in the protoplanetary disk the stochastic planetesimal
motions are of low velocity so that when they collide, they collide
essentially at velocities that are only just above their mutual escape
speed. Since much of the energy is dissipated in the inelastic impact, most
of the debris will not have enough kinetic energy to escape the system and
will (after perhaps a flyby or two) eventually accrete onto a single body or
a binary body. However, there is evidence that inside a single chondrite
there exist chondrules of different ages (Villeneuve et al.\ 2009). And some
chondrules experienced multiple heating events. Somehow chondrules produced
in multiple chondrule-forming impacts must have mixed together and form a
single chunk of rock that later is observed as a chondritic meteorite. The
immediate re-accretion scenario may make this somewhat difficult because
once the collision releases most of the magma, and much of this is
reaccreted as chondrules (for the molten part) and matrix (for the
non-molten part) the remaining body will no longer be a sphere of magma.

If, however, the collisions would be high-speed enough that much of the
debris escapes, the newly formed chondrules will disperse into the
protoplanetary disk. There they mix with the pool of other chondrules of
various ages, and then later accrete into a chondrite parent body.  An
important question in this scenario is how aerodynamic drag and the
resulting radial drift affects this: will the chondrules stay long enough
in the asteroid-belt-region before they dift away or not? Jacquet et al.
(2012) study this question in detail.

Perhaps more likely is a combination of the two. In particular the 
cluster chondrites of Metzler (2012, see Section \ref{sec-comp-chond})
can be easily understood in terms of the reaccretion scenario while
the multiple ages seems more to point toward a dispersion and later
accretion scenario.

\subsection{Volatile elements}\label{sec-volatile-elements}
One of the main characteristics of most chondrules is their ``normal''
  abundance of volatile elements such as Na, K etc. If a chondrule would
  stay at a high temperature (e.g.~2000K) for more than a few minutes, these
  elements would have evaporated out of the chondrule and leave a
  volatile-depleted chondrule behind. We believe that the impact splash
  scenario may naturally resolve this issue, because during the hot phases
  of the expanding impact splash the density of the cloud is so high that
  the vapor quickly reaches the saturation pressure and evaporation
  stops. However, this idea must be verified by actual modeling: Will the
  evaporation indeed saturate? Will the temperature drop below the
  evaporation temperature before the density of the expanding cloud becomes
  too low to saturate the pressure? Answering these questions with explicit
  calculations is the topic of paper II in this series.

\subsection{What kind of chondrules qualify for this scenario?}
So far we have not made any distinction between different kinds of
  chondrules or chondrites. Given that chondrule properties vary
  considerably between chondritic classes, the natural question arises: do
  all chondrules form through the same mechanism or are different chondrule
  formation scenarios responsible for different kinds of chondrules? And 
  for which kind, if at all, could the impact splash scenario of this paper
  be applicable?

A special class of chondrites and chondrules is the CB and CH class.
  It has been recognized for some time that these chondrules may indeed have
  originated from colliding planetesimals or planets (see Desch et al.~2012
  for a discussion). The evidence for this includes (1) the
  cryptocrystalline structure of CB chondrules, (2) their strong depletion
  of volatile elements, (3) zoned metal spherules that appear to have been
  condensed out of the gas phase in a highly energetic single-staged process
  (Krot et al. 2005) and (4) their young age, 5 Myr after CAIs. One can
  interpret these properties as meaning (1) that these chondrules must have
  cooled more rapidly (seconds to minutes) than ``normal'' chondrules
  (hours) to explain their textures, (2) that the vapor-melt plume of the
  impact must have rapidly become of low enough density to allow volatile
  elements to get out of the chondrules and not recondense, (3) the impact
  must have been of very high velocity (several km/s) to produce metal
  vapor, (4) which appears consistent with the young age which means low gas
  densities in the protoplanetary disk and thus allows for high-speed
  impacts. Arguments (1) and (2) seem to require a rapid reduction of the
  density and optical depth, which can be achieved by a high impact velocity
  (also required for (3) and consistent with (4)) and a low mass of the
  impactor. Krot et al.~(2005) argue, instead, for a very high mass
  collision to explain CB chondrules.  Perhaps the adiabatic cooling of the
  impact-shock-compressed rock is enough to explain the fast cooling, or
  only a fraction of the mass from the giant impact actually gets converted
  into a plume of chondrules.

In constrast to the case for the CB/CH chondrules our model may explain
  the properties of ``normal'' chondrules: (1) barred or porphyritic
  textures indicating slower cooling (hours), (2) little depletion of
  volatiles (see Section \ref{sec-volatile-elements}), (3) possibly lower
  temperatures and (4) earlier times after CAIs. The slower cooling rates
  come naturally out of our model because the optical depth of the cloud
  keeps the chondrules warm for a while. The little depletion of volatiles
  may be achieved by the high densities (though a quantitative analysis has
  to wait until paper II in this series). And the lower temperatures are
  natural for lower impact velocities which are expected in the earlier
  phases of the protoplanetary disk. The melt is the molten interior of
  the $^{26}$Al-heated planetesimals, which also is expected during the
  earlier phases.

We might also be able to apply our radiatively cooling expanding cloud
  model to the high impact velocity scenario for CB/CH chondrules, but it
  would require us to include more physics, in particular a proper equation
  of state for the shock-heated rock and the subsequent adiabatic expansion
  and adiabatic cooling.

Among the ``normal'' chondrules there is also the issue of radial and
  barred textures versus porphyritic textures. The latter are the majority
  (about 85\% of chondrules, Gooding \& Keil 1981). Radial textures require
  that no nucleation sites were present in the melt, so that crystallization
  in the supercooled melt droplet starts from a single site and progresses
  radially outward from that site. In the splash scenario this means that
  the melt in the original pre-molten planetesimals was well above the
  liquidus, so that any previously existing crystals were molten.
  Conversely, for porphyritic chondrules numerous nucleation sites must have
  been present in the melt, and thus must have presumably remained present
  in the magma of the colliding planetesimals, meaning that that magma was
  likely slightly below the liquidus temperature. The impact may have heated
  the melt through shock-heating above the liquidus, but if the cooling sets
  in quickly enough some of these nucleation sites may survive. The question
  of porphyritic textures and the origin of the nucleation sites in the
  splash scenario remains, however, a tricky one. Sanders \& Scott (2012)
  propose instead that these nucleation sites may have entered into the melt
  droplets as pollution from e.g.\ the surface regolith that may have been
  mixed in with the chondrules. The problem with this (as with the relict
  grain issue of Section \ref{sec-relict-grains}) is that the timing of
  mixing must be ideal: too early and the temperature is still above the
  liquidus and these regolith particles will also melt; too late and the
  chondrules have already formed radial or barred textures.

Finally there is the issue of Fe-content of chondrules: the H, L and LL
  classes. If the pre-molten planetesimals have experienced strong
  differentiation, it is possible that the mantles of these planetesimals
  are poor in iron and nickel. If the masses of the planetesimals are low
  enough and convection might be present, then perhaps all too strong loss
  of Fe in the mantle can be avoided. The issue of H, L and LL chondrites
  in the context of the splash scenario is further discussed in Sanders
  \& Scott (2012).

\section{Summary and conclusion}
This paper studies the radiative cooling of an expanding cloud of lava
droplets originating from the low-velocity collision between two partly or
fully pre-molten planetesimals. The model is also valid for high-speed
collisions between cold planetesimals, in which the melt is produced by the
collision itself.

The main result of this paper is a self-consistent cooling profile for
chondrules based on the initial conditions of the expanding plume of melt
droplets (chondrules). These cooling profiles are characterized by a
time scale $t_{\mathrm{cool}}$ given by Eq.~(\ref{eq-tcool}) and are well
approximated by a constant temperature for $t<t_{\mathrm{cool}}$ and a
powerlaw dropoff (Eq.~\ref{eq-t-analytic}) for $t>t_{\mathrm{cool}}$. The
cooling rate is given by Eq.~(\ref{eq-temp-cool-analyt}). The cooling time
scale $t_{\mathrm{cool}}$ depends on the total mass of the plume and on its
expansion velocity.  Both parameters are related to the conditions of the
collision between the two planetesimals.

A large total mass of the expanding cloud of chondrules means that the
colliding planetesimals must have been at least of that mass, but presumably
considerably more massive since not all of the matter is likely to end up in
a single expanding plume. The expansion velocity of the droplet cloud is
related to the collision velocity of the two planetesimals: it is unlikely
that it is higher than the impact velocity. Note that the expansion velocity
is different from (and presumably much less than) the velocity by which the
cloud moves away from the impact site. To relate both parameters
($M_{\mathrm{cloud}}$ and $v_{\mathrm{exp}}$) to paramters of the colliding
planetesimals it is necessary to perform 3-D hydrodynamic simulations of the
impact, such as those performed by Asphaug et al.~(2011).

The results of our model, when compared to known cooling time scales of
chondrules, put constraints on $M_{\mathrm{cloud}}$ and $v_{\mathrm{exp}}$,
and thus on the conditions under which planetesimal collisions can 
produce chondrules.

{\bf Acknowledgements:} We would like to thank Andreas Pack, Knut Metzler,
Alessandro Morbidelli, David Lundberg, Addi Bischoff, Neal Turner,
Stephan Henke and Mark Swain for useful discussions and feedback.
This paper benefitted a lot from detailed comments and suggestions
by an anonymous referee, for which we are very grateful.
S.\ Stammler is funded by the Deutsche Forschungsgemeinschaft (DFG)
grant Du 414/12-1.

\appendix

\section{Time-dependent radiative transfer: Discretization and implicit
  integration}
\label{app-rt-eq-discrete}
Solving the radiative transfer moment equations coupled to the
time-dependent heating/cooling equation requires an implicit
integration scheme. The two coupled equations to solve are:
\begin{equation}
F_1(J,T):=\frac{1}{\alpha}\frac{1}{r^2}\frac{d}{dr}\left\{
\frac{r^2}{\alpha}\left[\frac{d(fJ)}{dr}+\frac{3f-1}{r}J\right]
\right\}-J+\frac{\sigma}{\pi}T^4
\end{equation}
and
\begin{equation}
F_2(J,T):=\rho c_v\frac{dT}{dt}-4\pi\alpha\left(J-\frac{\sigma}{\pi}T^4\right)=0
\end{equation}
The inner boundary condition is:
\begin{equation}
F_1(J,T,r=r_{\mathrm{in}}):= H(r=r_{\mathrm{in}}) = 0
\end{equation}
where $r_{\mathrm{in}}$ is the smallest radius of our grid, which we take
$r_{\mathrm{in}}\ll R_{\mathrm{cloud}}$. This condition translates into
\begin{equation}
\left[\frac{d(fJ)}{dr}+\frac{3f-1}{r}J\right]_{r=r_{\mathrm{in}}} = 0
\end{equation}
The outer boundary condition is 
\begin{equation}
F_1(J,T,r=r_{\mathrm{out}}):=H(r=r_{\mathrm{out}}) - hJ = 0
\end{equation}
with $h=H_{\mathrm{fte}}(r_{\mathrm{out}})/J_{\mathrm{fte}}(r_{\mathrm{out}})$,
which translates into
\begin{equation}
F_1(J,T,r=r_{\mathrm{out}}) := \left[\frac{d(fJ)}{dr}+\frac{3f-1}{r}J+h\alpha J\right]_{r=r_{\mathrm{out}}} = 0
\end{equation}
The only time-derivative in the equations is the one on the temperature.
The radiation field is assumed to immediately adapt. 

We put $J(r,t)$ and $T(r,t)$ on a spatial grid $\{r_1,\cdots,r_N\}$. So for
time step $n$ we have the values $J^n_1,\cdots,J^n_N$ and
$T^{n}_1,\cdots,T^{n}_N$. The above equations are also to be evaluated at
these grid points: $F_{1,1}=0,\cdots,F_{1,N}=0$ and
$F_{2,1}=0,\cdots,F_{2,N}=0$. Let us define
\begin{equation}
Q^n=\left(J_1^n,T_1^{n},J_2^n,T_2^{n},\cdots,J_N^n,T_N^{n}\right)^T
\end{equation}
and
\begin{equation}
F(Q)=\left(F_{1,1}(Q),F_{2,1}(Q),F_{1,2}(Q),F_{2,2}(Q),\cdots,F_{1,N}(Q),F_{2,N}(Q)\right)^T
\end{equation}
The objective of the time-integration of these equations is to find the
values of $Q^{n+1}$: the values of $Q$ at the next time step. 

For numerical stability we express all instances of $Q$ in the equations $F$ 
in their future form: $Q^{n+1}$, except when a time-derivative of $Q$ is used, which
is written as $\partial Q/\partial t=(Q^{n+1}-Q^{n})/\Delta t$ and where the
present value of $Q^{n}$ is required. We thus get as our set of equations:
\begin{equation}
F(Q^{n+1})=0
\end{equation}
which we need to solve for $Q^{n+1}$. If $F$ were linear in $Q^{n+1}$ then
one could write the vector $F$ as a matrix multiplication with the vector
$Q^{n+1}$ plus a contant vector $R$:
\begin{equation}
F=MQ^{n+1}+R
\end{equation}
Then the solution to $F=0$ would be just a matrix inversion:
\begin{equation}
Q^{n+1}=-M^{inv}R
\end{equation}
However, $F$ is non-linear in $Q^{n+1}$ because the Planck function
$B(T)=(\sigma/\pi)(T^{n+1})^4$. And so we have a non-linear function
$F(Q^{n+1})$.  So let us start with an initial guess $Q^{n+1}_{(0)}$. We
typically then have
\begin{equation}
F(Q^{n+1}_{(0)})\neq 0
\end{equation}
Ideally we want to find a $Q^{n+1}_{(1)}$ for which $F(Q^{n+1}_{(1)})=0$, but we
can only use Newton's method by approximating $F(Q^{n+1}_{(1)})$ using 
first order Taylor expansion:
\begin{equation}
F(Q^{n+1}_{(1)})\simeq F(Q^{n+1}_{(0)})+\frac{\partial F}{\partial Q}\cdot (Q^{n+1}_{(1)}-Q^{n+1}_{(0)}) =0
\end{equation}
or more in general for the $k$-th iteration:
\begin{equation}
F(Q^{n+1}_{(k+1)})\simeq F(Q^{n+1}_{(k)})+\frac{\partial F}{\partial Q}\cdot (Q^{n+1}_{(k+1)}-Q^{n+1}_{(k)}) =0
\end{equation}
If we define
\begin{equation}
\Delta Q^{n+1}_{(k)}=Q^{n+1}_{(k+1)}-Q^{n+1}_{(k)}
\end{equation}
we can solve for $\Delta Q^{n+1}_{(k)}$:
\begin{equation}
\Delta Q^{n+1}_{(k)} = -\left(\frac{\partial F}{\partial Q}\right)^{inv}F(Q^{n+1}_{(k)})
\end{equation}
and then we obtain the new:
\begin{equation}
Q^{n+1}_{(k+1)}=\Delta Q^{n+1}_{(k)}+Q^{n+1}_{(k)}
\end{equation}

Now let us write out the discrete equations $F_1(J,T)$ and $F_2(J,T)$ explicitly:
\begin{equation}
\begin{split}
F_{1,i}(J,T^{n+1}) \equiv  & \frac{1}{\alpha_ir_{i}^2\Delta r_{i}}\bigg(
\frac{r_{i+1/2}^2}{\alpha_{i+1/2}}\frac{f_{i+1}J_{i+1}-f_{i}J_{i}}{\Delta r_{i+1/2}} \\
& \qquad\qquad 
-\frac{r_{i-1/2}^2}{\alpha_{i-1/2}}\frac{f_{i}J_{i}-f_{i-1}J_{i-1}}{\Delta r_{i-1/2}}\\
& \qquad\qquad 
+\frac{(3f_{i+1/2}-1)(J_{i+1}+J_i)r_{i+1/2}}{2\alpha_{i+1/2}}\\
& \qquad\qquad 
-\frac{(3f_{i-1/2}-1)(J_{i-1}+J_i)r_{i-1/2}}{2\alpha_{i-1/2}}\bigg)\\
& \quad -J_i+\frac{\sigma}{\pi}(T_i^{n+1})^4 = 0
\end{split}
\end{equation}
For the gas equation we get:
\begin{equation}
F_{2,i}(J,T^{n+1})\equiv 
\rho c_v\frac{T_i^{n+1}-T_i^n}{\Delta t}-
4\pi\alpha_i\left(J_i-\frac{\sigma}{\pi}(T_i^{n+1})^4\right)
-q=0
\end{equation}
Boundary conditions only have to be applied to the first equation. 
The inner boundary condition:
\begin{equation}
F_{1,1}(J,T^{n+1})=\frac{f_2J_2-f_1J_1}{\alpha_{3/2}\Delta r_{3/2}}
+\frac{1}{2}(J_1+J_2)\frac{3f_{3/2}-1}{\alpha_{3/2}r_{3/2}}=0
\end{equation}
The outer boundary condition:
\begin{equation}
F_{1,N}(J,T^{n+1})=\frac{f_{N}J_{N}-f_{N-1}J_{N-1}}{\alpha_{N-1/2}\Delta r_{N-1/2}}+
\frac{1}{2}(J_{N}+J_{N-1})\left(
\frac{3f_{N-1/2}-1}{\alpha_{N-1/2}r_{N-1/2}}+h\right)=0
\end{equation}
The energy equation can be rescaled to:
\begin{equation}
F_{2,i}(J,T^{n+1})\equiv 
T_i^{n+1}-T_i^n-
\frac{4\pi\alpha_i\Delta t}{\rho c_v}\left(J_i-\frac{\sigma}{\pi}(T_i^{n+1})^4\right)
-\frac{q\Delta t}{\rho c_v}
=0
\end{equation}
The matrix coefficients then become:
\begin{eqnarray}
\left(\frac{\partial F_{1,i}}{\partial J_i}\right) &=& 
-\frac{1}{\alpha_ir_i^2\Delta r_{i}}\left\{
\frac{r_{i-1/2}^2f_i}{\alpha_{i-1/2}\Delta r_{i-1/2}}
+\frac{r_{i+1/2}^2f_i}{\alpha_{i+1/2}\Delta r_{i+1/2}}\right\}
-1\\
\left(\frac{\partial F_{1,i}}{\partial J_{i+1}}\right) &=&
\frac{1}{\alpha_ir_i^2\Delta r_i}\left\{\frac{r_{i+1/2}^2f_{i+1}}{\alpha_{i+1/2}\Delta r_{i+1/2}}
+\frac{(3f_{i+1/2}-1)r_{i+1/2}}{2\alpha_{i+1/2}}\right\}\\
\left(\frac{\partial F_{1,i}}{\partial J_{i-1}}\right) &=&
\frac{1}{\alpha_ir_i^2\Delta r_i}\left\{\frac{r_{i-1/2}^2f_{i-1}}{\alpha_{i-1/2}\Delta r_{i-1/2}}
+\frac{(3f_{i-1/2}-1)r_{i-1/2}}{2\alpha_{i-1/2}}\right\}\\
\left(\frac{\partial F_{1,i}}{\partial T_{i}^{n+1}}\right) &=&
\frac{4\sigma}{\pi}(T^{n+1}_i)^3\\
\left(\frac{\partial F_{2,i}}{\partial J_i}\right) &=& -\frac{4\pi\alpha_i}{\rho_ic_v}\Delta t\\
\left(\frac{\partial F_{2,i}}{\partial T_{i}^{n+1}}\right) &=&
1+\frac{4\pi\alpha_i}{\rho_i c_v}\frac{4\sigma}{\pi}(T^{n+1}_i)^3\Delta t
\end{eqnarray}
The matrix coefficients for the boundary conditions become:
\begin{eqnarray}
\left(\frac{\partial F_{1,1}}{\partial J_2}\right) &=& 
\frac{1}{\alpha_{3/2}}
\bigg\{\frac{3f_{3/2}-1}{2r_{3/2}}+\frac{f_2}{\Delta r_{3/2}}\bigg\}\\
\left(\frac{\partial F_{1,1}}{\partial J_1}\right) &=& 
\frac{1}{\alpha_{3/2}}
\bigg\{\frac{3f_{3/2}-1}{2r_{3/2}}-\frac{f_1}{\Delta r_{3/2}}\bigg\}\\
\left(\frac{\partial F_{N,1}}{\partial J_{N}}\right) &=& 
\frac{1}{\alpha_{N-1/2}}\left(
\frac{3f_{N-1/2}-1}{2r_{N-1/2}}+\frac{f_N}{\Delta r_{N-1/2}}\right)+\frac{h}{2}\\
\left(\frac{\partial F_{N,1}}{\partial J_{N-1}}\right) &=& 
\frac{1}{\alpha_{N-1/2}}\left(
\frac{3f_{N-1/2}-1}{2r_{N-1/2}}-\frac{f_{N-1}}{\Delta r_{N-1/2}}\right)+\frac{h}{2}
\end{eqnarray}

\end{document}